\documentclass[%
 aip,
 amsmath,amssymb,
 reprint,%
]{revtex4-1}

\usepackage{graphicx}
\usepackage{subfigure}
\usepackage{float}
\usepackage{dcolumn}
\usepackage{bm}
\usepackage{textcomp}

\usepackage[utf8]{inputenc}
\usepackage[T1]{fontenc}
\usepackage{mathptmx}

\begin{document}

\preprint{AIP/123-QED}

\title{Free-standing silicon nitride nanobeams with efficient fiber-chip interface for cavity QED}

\author{Abdulrahman Alajlan}
 \affiliation {Texas A\&M University, Department of Physics and Astronomy, 4242 TAMU, College Station, Texas, USA}
 \affiliation {King Abdulaziz City for Science and Technology, Riyadh 11442, Saudi Arabia}
 
\author{Mohit Khurana}
 \affiliation {Texas A\&M University, Department of Physics and Astronomy, 4242 TAMU, College Station, Texas, USA}

\author{Xiaohan Liu}
 \affiliation {Texas A\&M University, Department of Physics and Astronomy, 4242 TAMU, College Station, Texas, USA}
 
\author{Ivan Cojocaru}
 \affiliation {Texas A\&M University, Department of Physics and Astronomy, 4242 TAMU, College Station, Texas, USA}
 \affiliation {Russian Quantum Center, 100A, Novaya Street, Skolkovo, 143025 Moscow, Russia}
 \affiliation {PN Lebedev Institute RAS, Leninsky Prospect 53, 119991 Moscow, Russia}
 
\author{Alexey V. Akimov}
 \affiliation {Texas A\&M University, Department of Physics and Astronomy, 4242 TAMU, College Station, Texas, USA}
 \affiliation {Russian Quantum Center, 100A, Novaya Street, Skolkovo, 143025 Moscow, Russia}
 \affiliation {PN Lebedev Institute RAS, Leninsky Prospect 53, 119991 Moscow, Russia}

\date{\today}

\begin{abstract}
We present the design, fabrication and characterization of high quality factor silicon nitride nanobeam PhC cavities at visible wavelengths for coupling to diamond color centers in a cavity QED system. We demonstrate devices with a quality factor of $\sim 24,000$ around the zero-phonon line of the germanium-vacancy center in diamond. We also present an efficient fiber-to-waveguide coupling platform for suspended nanophotonics. By gently changing the corresponding effective indices at the fiber-waveguide interface, we achieve an efficiency of $\sim 96\%$ at the cavity resonance.

\end{abstract}

\maketitle

The recent progress in quantum technologies has increased the demands for a robust scalable platform to realize an efficient quantum interface between light and matter. Building such interface is important for many applications ranging from optical sensing and metrology \cite{Q_metrolgy_sening_book} to quantum information and quantum computation \cite{Q_Computation_book_Nielsen_chuang}. One of the most promising type of platforms involves an atom coupled to optical resonator to allow a coherent energy exchange in a cavity QED system \cite{Cirac_Qtransfere, QD_QED_Imamog, Single_Ph_transistor}. Towards this goal, several micro-cavities have been developed including Fabry-Perot cavities \cite{fabry_perot_kimble,Fabry-finesse_PRL_2007}, silica microspheres \cite{microsphere_Kimble_1998}, and microtoroids \cite{Micro_toroid_Vahala_2003}. Despite the fact that aforementioned cavities possess ultra-high finesses and quality factors, their relatively large volume makes the integration of many devices in one chip very challenging. Instead, photonic crystal (PhC) cavities  have recently emerged as a scalable platform for efficient light-matter interactions due to their small mode volumes accompanied with high quality factors \cite{QD_Cavity_Imamogulu, faraon_NV_diamond_cavity_coupling,Lukin_SiV_Cavity_Science,MoS2_Cavity_QED,Carbon_nanotube_cavity}. 

In this framework, Silicon nitride offers many advantages over other alternative materials, mostly due to its wide band gap and compatibility with standard CMOS processes. In visible wavelengths, however, the moderately low index of refraction has limited the quality factor of PhC cavities  on Si\textsubscript{3}N\textsubscript{4} films to only 3 orders of magnitude in visible wavelengths \cite{SiN_Q_factor_1,SiN_Q_factor_2,SiN_Q_factor_3,SiN_Q_factor_4,SiN_Q_factor_5}. In this report, we demonstrate nanobeam PhC cavities with a quality factor of $\sim 24,000$, that is 1 order of magnitude higher than previously reported.

On the other hand, nanobeam PhC cavities are not inherently suited to free-space optical coupling or fiber coupling due to the mismatch in the mode size and the effective refractive index. Previous work have used either notches \cite{Diamond_notch_Loncar_2014} or inverse-designed vertical couplers \cite{inverse_design_coupler_Jelena_2019} for free-space interfaces, and end-fire technique for fiber interfaces \cite{end_fire_2013}. The most efficient technique is based on evanescent coupling from a suspended waveguide to a biconical fiber taper \cite{adiabatic_coupling_painter_2013} with an efficiency over 90$\%$. Here, we employ the same principle using a single-sided optical fiber tip and demonstrate a coupling efficiency of 96$\%$, similar to detached devices reported by Tiecke \textit{et al.}\cite{fiber_optica_2015}.  
\begin{figure}
\centering     
\subfigure[]{\label{fig_1_a}\includegraphics[width=0.2\textwidth]{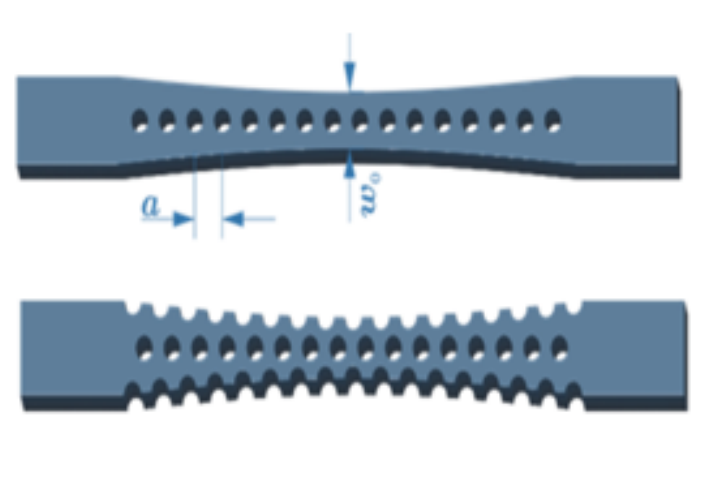}}
\subfigure[]{\label{fig_1_b}\includegraphics[width=0.2\textwidth]{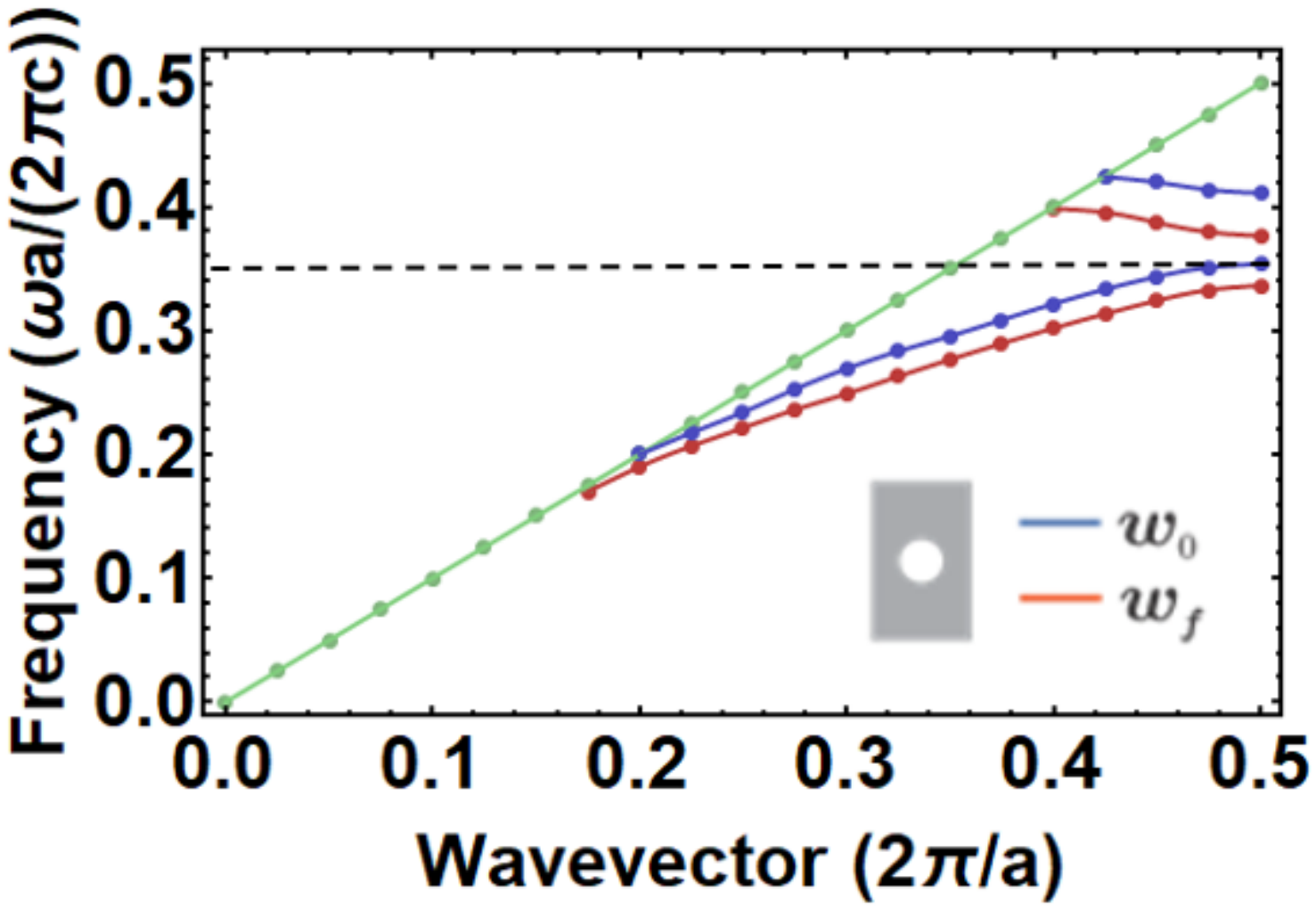}}
\subfigure[]{\label{fig_1_c}\includegraphics[width=0.2\textwidth]{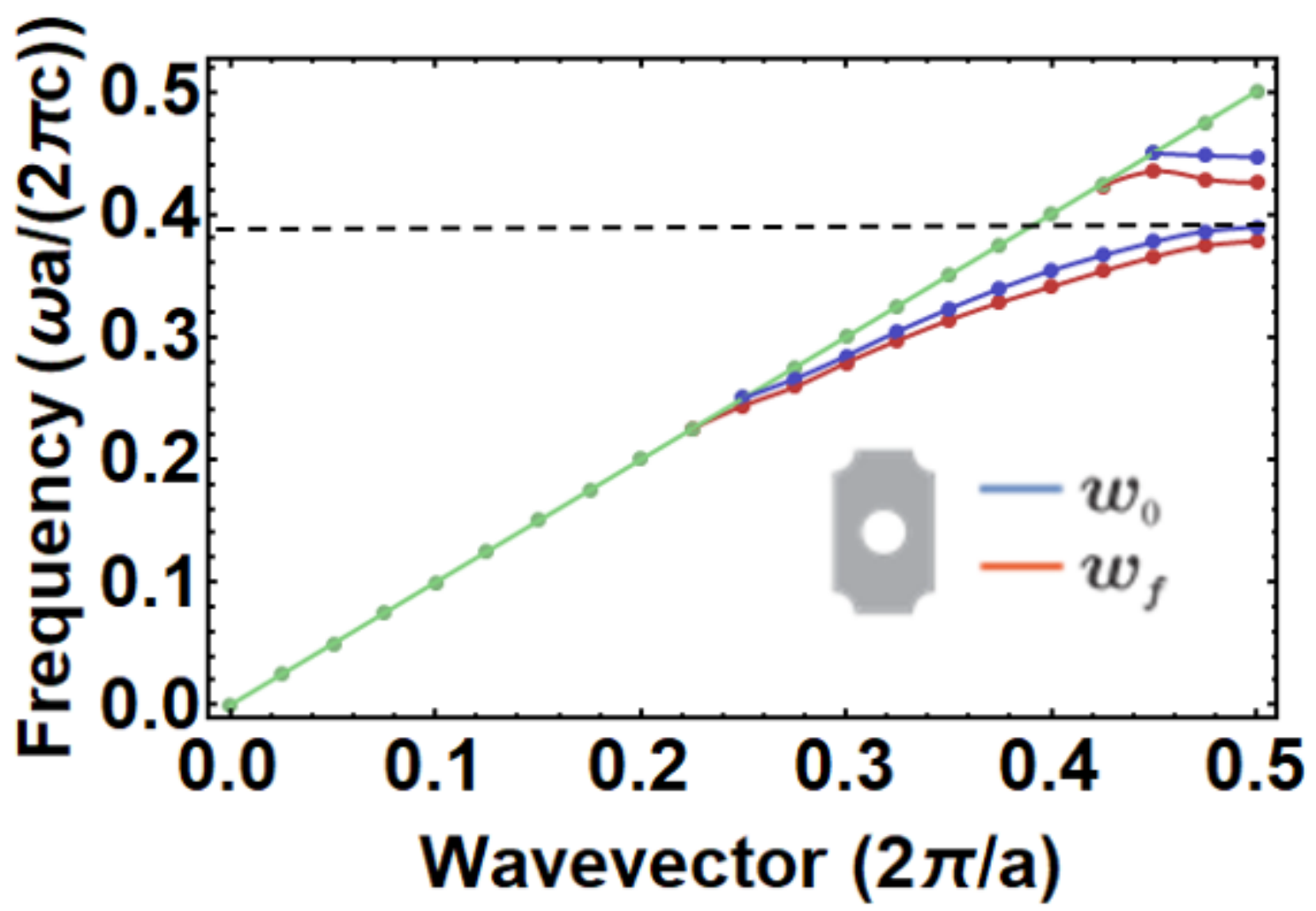}}
\subfigure[]{\label{fig_1_d}\includegraphics[width=0.2\textwidth]{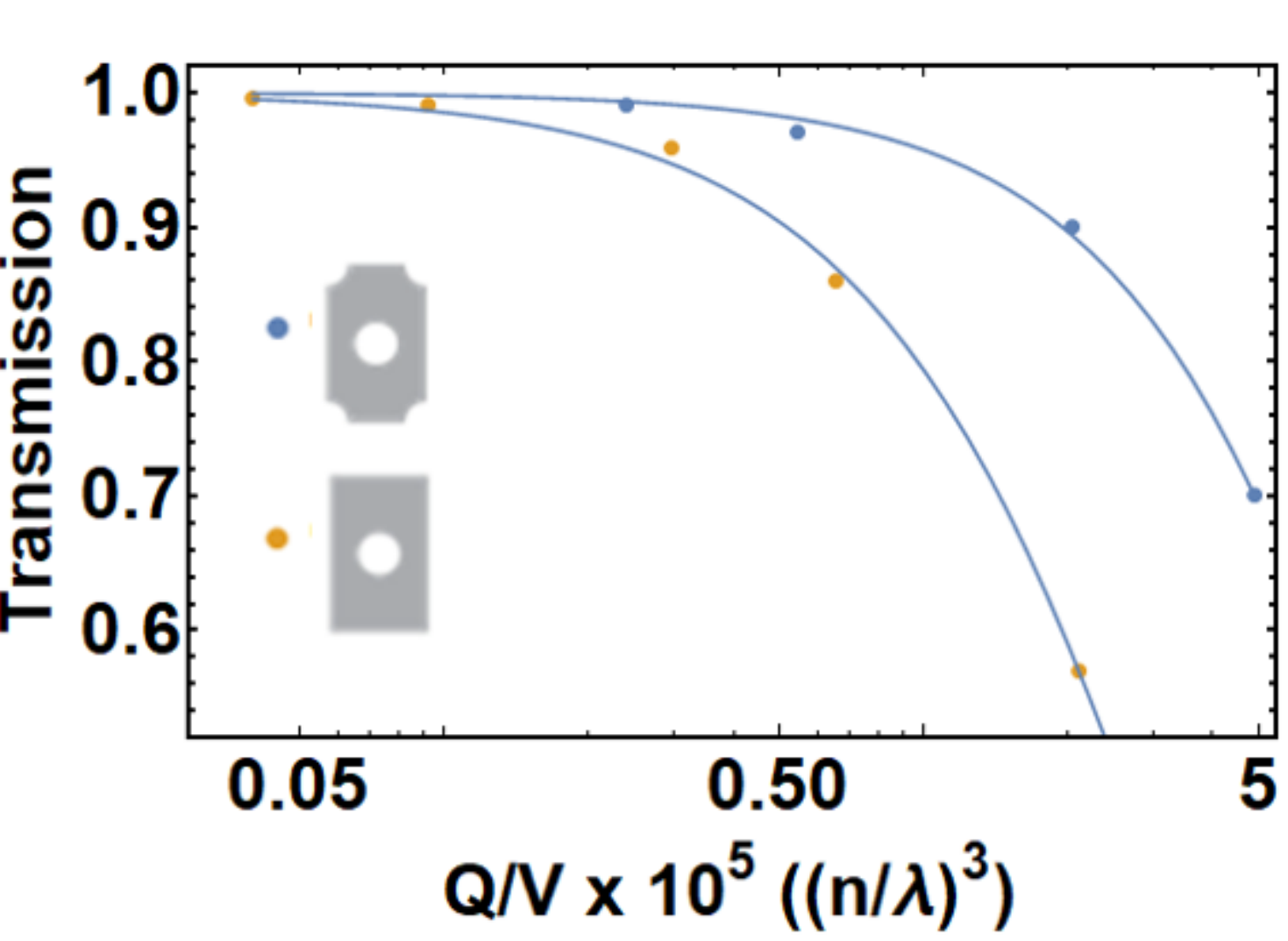}}
\caption{(a) Sketch of Si\textsubscript{3}N\textsubscript{4} nanobeam PhC cavity design adopted from Ahn \textit{et al.}\cite{Ahn_2010} (top) and from Alajlan \textit{et al.}\cite{Compact_design_2019} (bottom). (b) \& (c) The corresponding band structure for TE modes calculated by FDTD simulations. The dashed line marks the resonant frequency. (d) Transmission dependence on $Q/V$ value. The initial width is W\textsubscript{0}/a = 1.5, and the final width is W\textsubscript{f}/a = 2.}
\end{figure}

For the demonstration detailed here, we employed two designs (device A proposed by \cite{Ahn_2010} and device B proposed by Ahn \textit{et al.}\cite{Ahn_2010} and from Alajlan \textit{et al.}\cite{Compact_design_2019}) in which the nanobeam width is tapered smoothly while other parameters remain constant, as depicted in Fig. \ref {fig_1_a}. We note that both devices are designed based on the quadratic tapering method which is meant to generate a Gaussian-like field profile inside the cavity \cite{2D_High_Q_Akahane_2003,Nanobeam_OSA_Loncar}. In both devices, we focused only on TE modes since the TM bandgap is absent. The corresponding band diagrams, displayed in Fig. 1 (b) and (c), show that adding semicircular holes in device B shifts the resonant frequency slightly towards the light line resulting in extra radiation loss. Nevertheless, this can be mitigated by increasing the width of the cavity or reducing the holes radii. The parameters have been adjusted such that the simulated intrinsic quality factor for both devices reaches $\sim 10^6$. In practice, the spacing between holes needs to be adjusted to the point where the resonant frequency matches the target frequency. In our case, we adjusted the resonant frequency around $602$ nm, the zero-phonon line emission of the germanium-vacancy color center in diamond \cite{GeV_diamond_2015}. The cavity transmission and quality factor dependence is shown in Fig. \ref {fig_1_d}. It can be seen that the quality factor in device B exceeds the other one to some extent, similar to the case for gallium phosphide substrate \cite{Compact_design_2019}. The trade-off, however, is that device B is more sensitive to the fabrication imperfections which can be mitigated using a standard e-beam lithography (EBL) system explained next.

\begin{figure}
\centering     
\subfigure[]{\label{SEM_Far}\includegraphics[width=0.2015\textwidth]{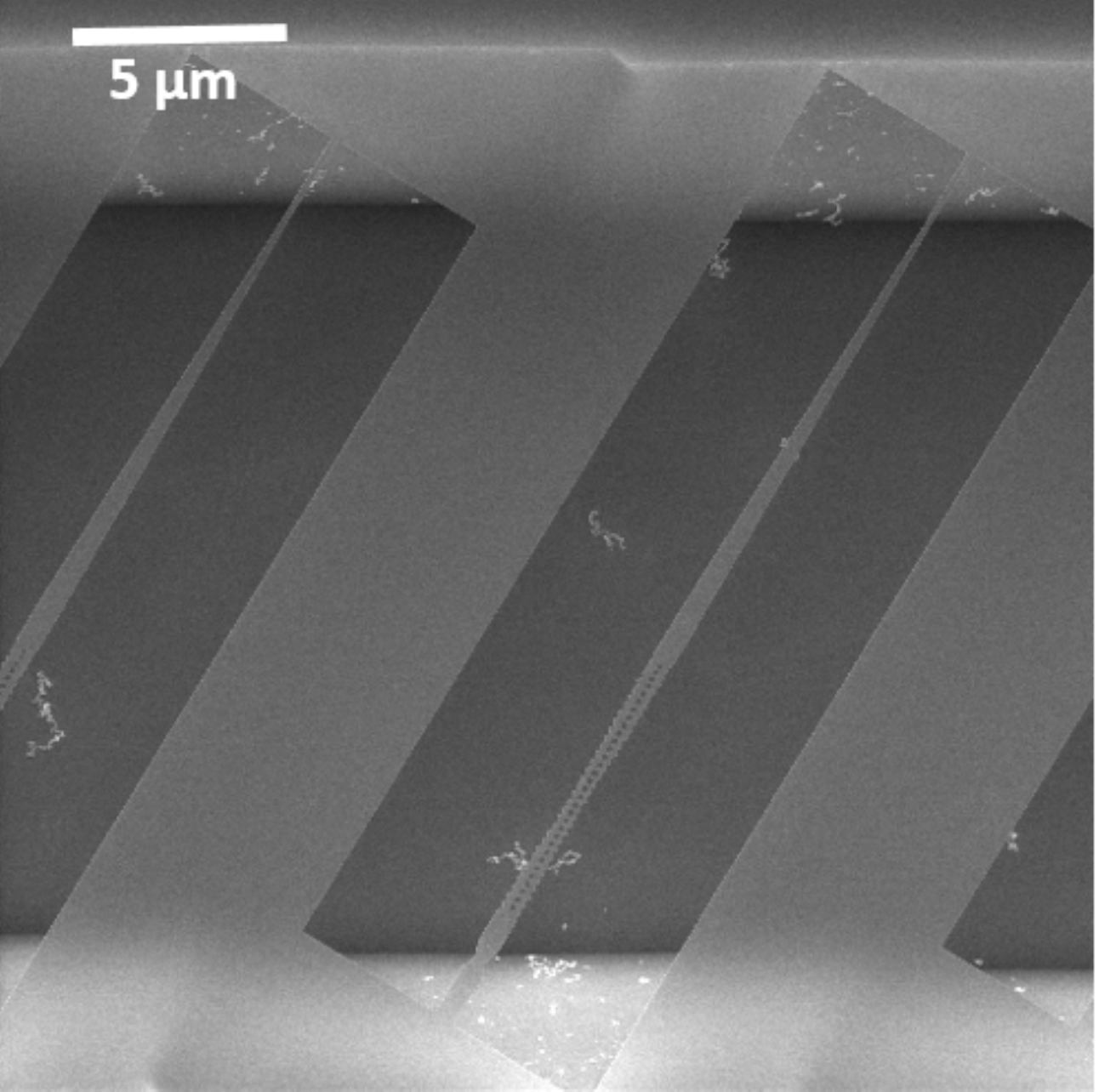}}
\subfigure[]{\label{SEM_Taper}\includegraphics[width=0.2\textwidth]{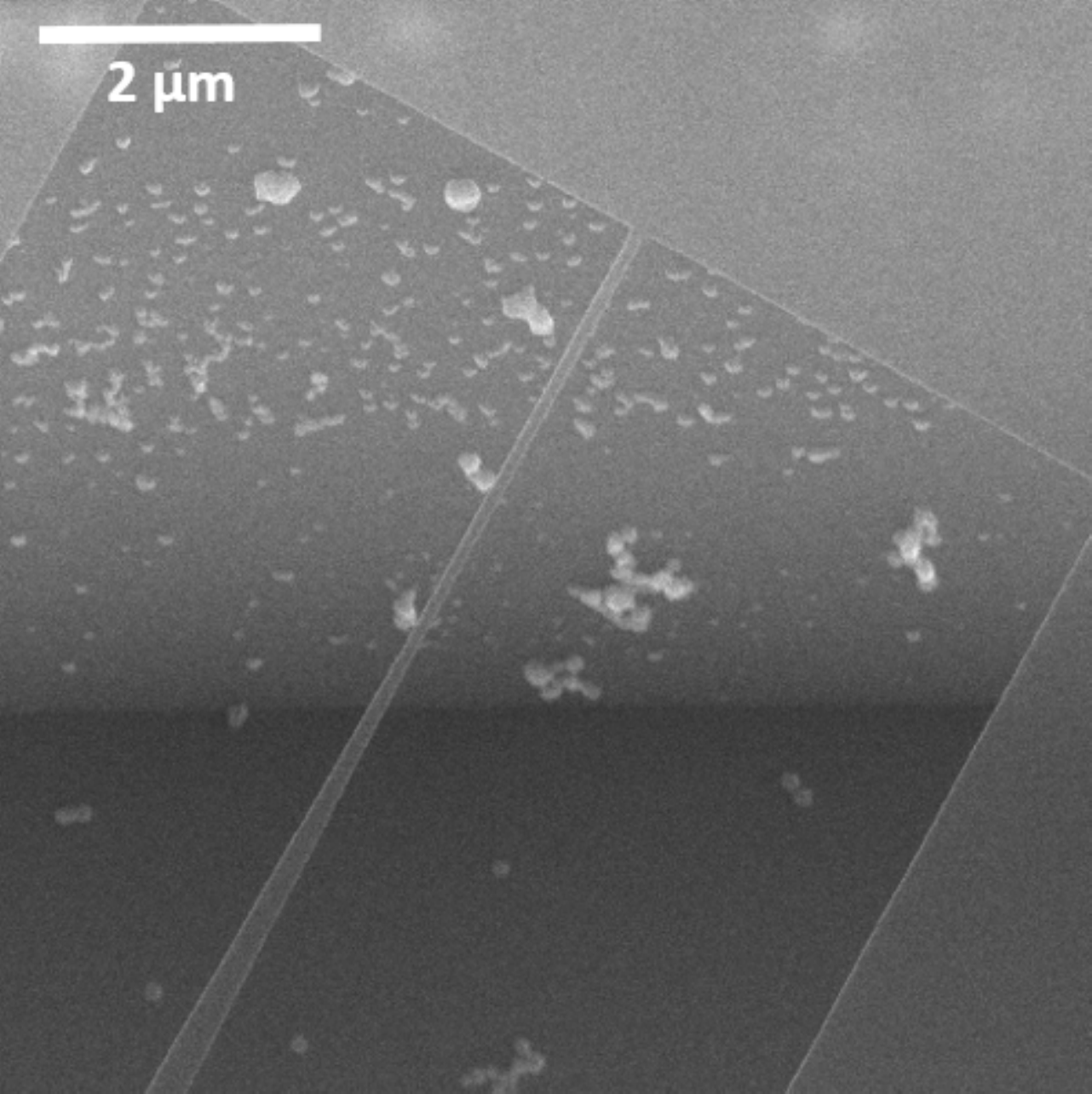}}
\subfigure[]{\label{SEM_Shift}\includegraphics[width=0.2015\textwidth]{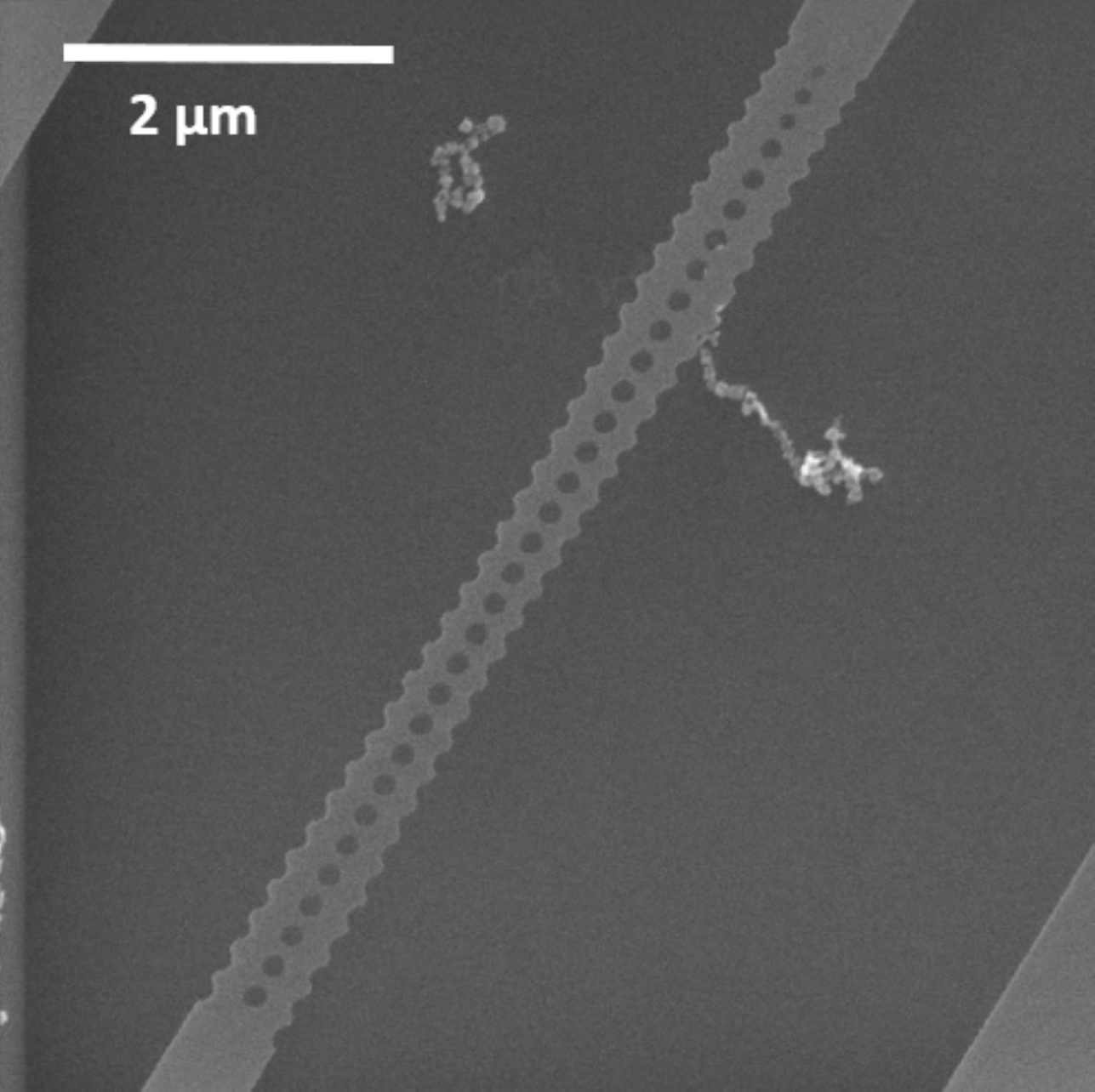}}
\subfigure[]{\label{SEM_Ahn}\includegraphics[width=0.2\textwidth]{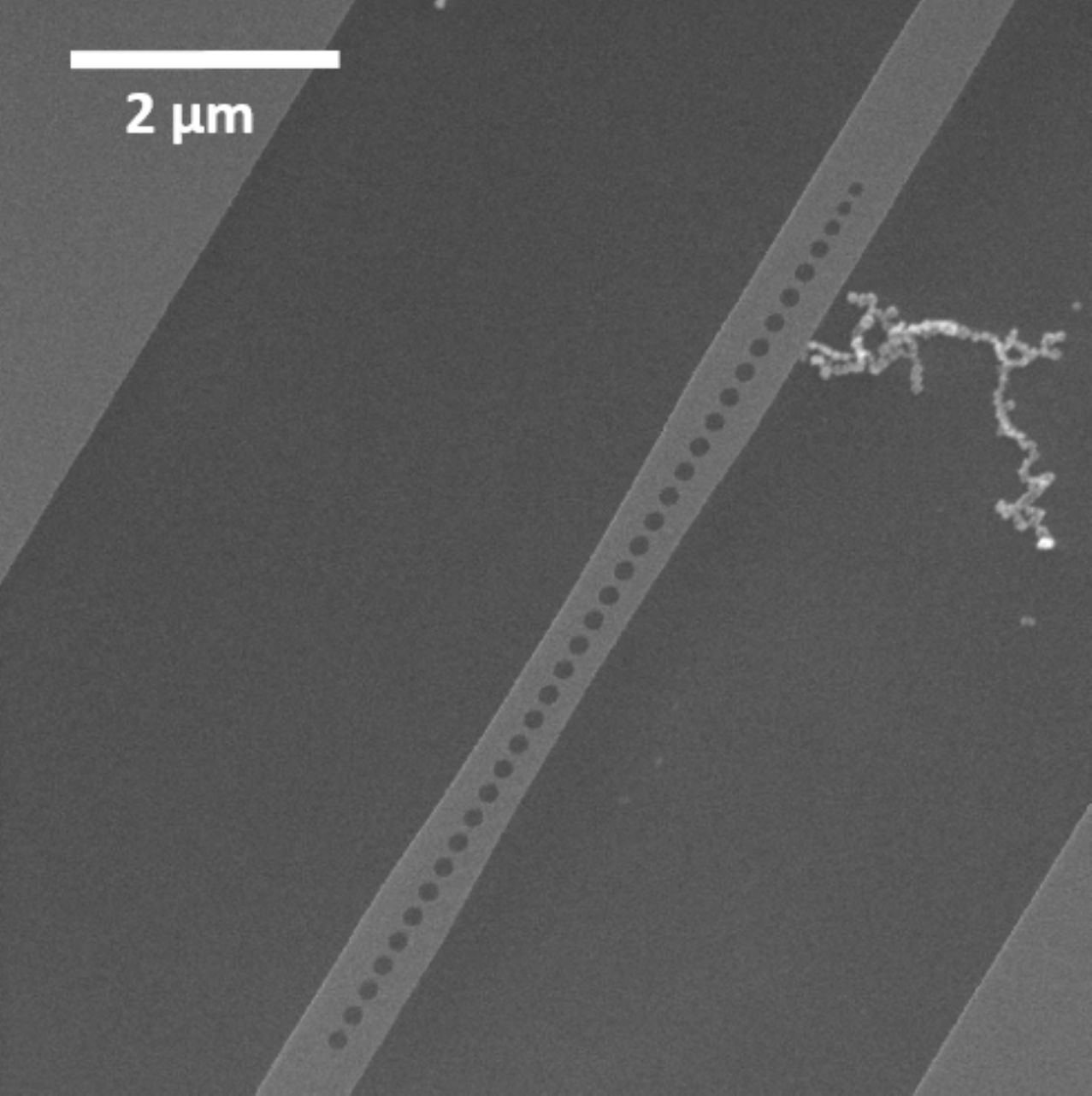}}
\caption{(a) SEM images of an array of Si\textsubscript{3}N\textsubscript{4} nanophotonic structures that consist of integrated nanobeam PhC cavities (c \& d) and free-standing waveguide tapered at the end (b).}
\label{SEM_Images}
\end{figure}

We carried out the fabrication on 200 nm thick stoichiometric Si\textsubscript{3}N\textsubscript{4} membrane deposited via low-pressure chemical-vapor deposition on a plain Si substrate. Polymethyl methacrylate (PMMA) is first spun directly on top of the silicon nitride membrane without a metallic layer or conductive polymer \cite{Escaper_SiN_Si_microresonator}. The resist is then patterned using an EBL system (TESCAN MIRA3) with an accelerating voltage of 30 kV, before developed in MIBK developer mixed with Isopropyl solution 1:3. Following the EBL process, the pattern is transferred to the Si\textsubscript{3}N\textsubscript{4} using CHF\textsubscript{3}-based dry etching (ICP-RIE, Oxford Instruments). After stripping the resist, $30\%$ KOH solution is used to etch the exposed silicon at 100 \textdegree{}C for 10 minutes. SEM images of the final device is shown in Fig. \ref{SEM_Images}.  

\begin{figure}
\centering     
\subfigure[]{\label{laser_scan_setup}\includegraphics[width=0.21\textwidth]{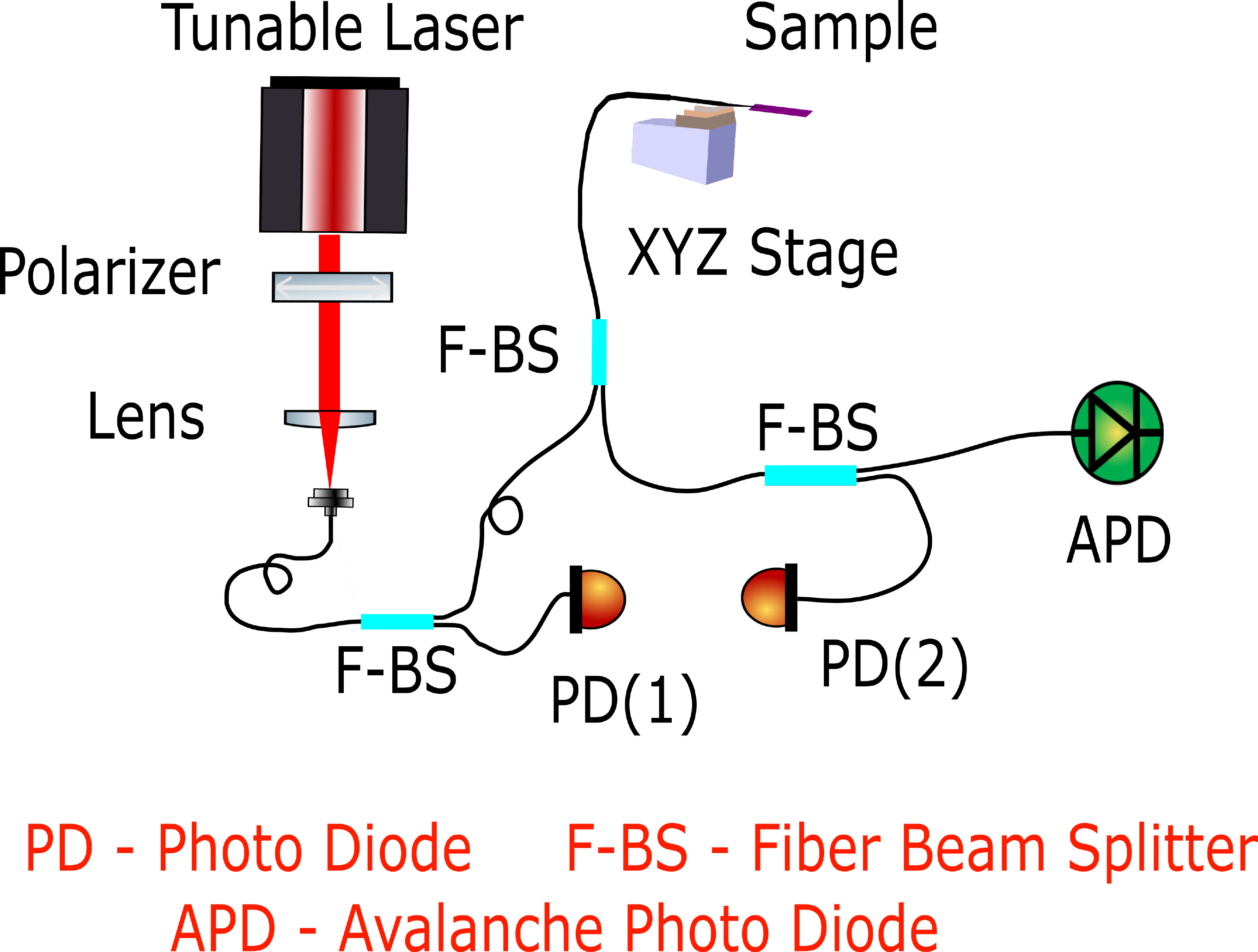}}
\subfigure[]{\label{laser_scan}\includegraphics[width=0.21\textwidth]{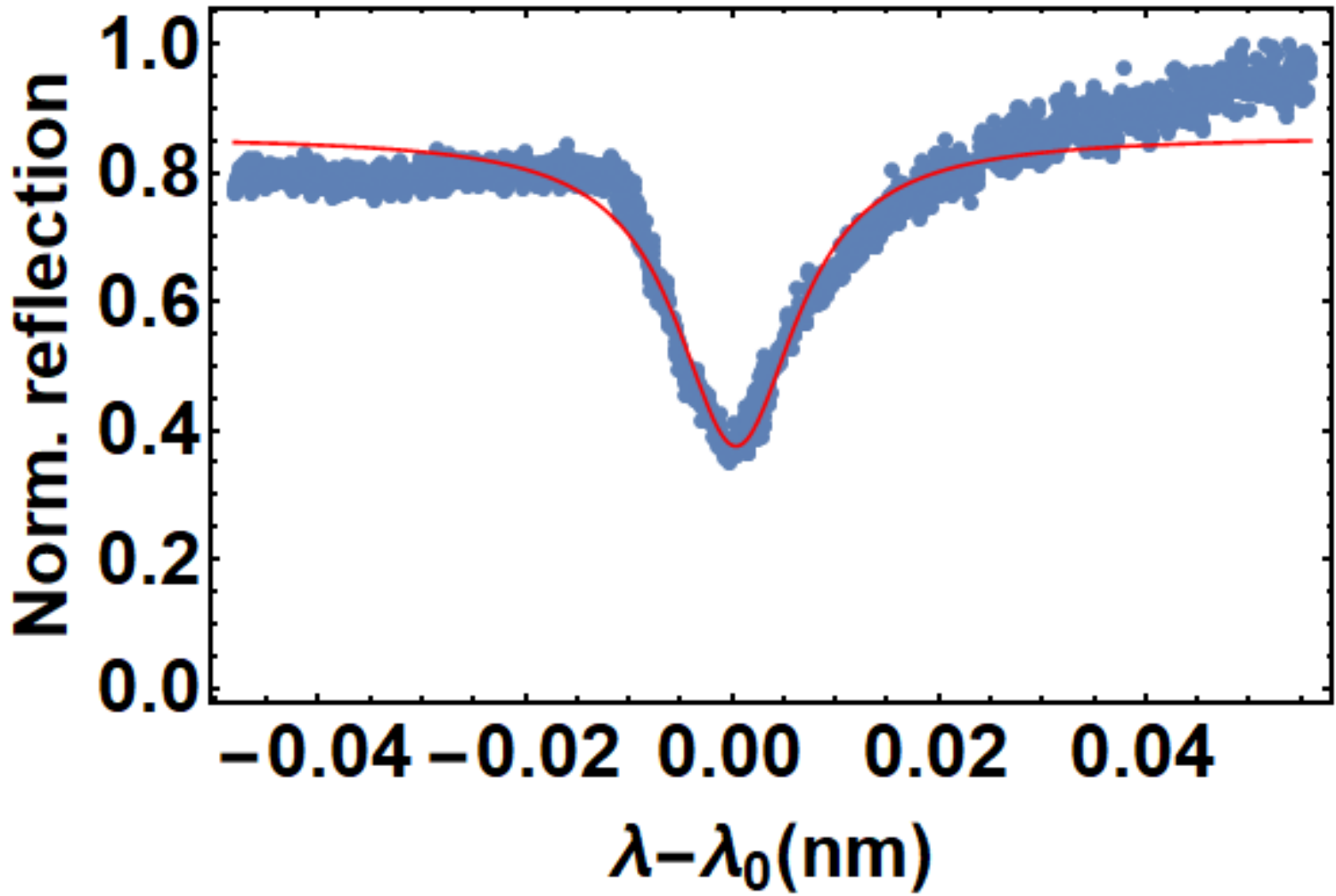}}

\caption{(a) Schematic optical characterization of Si\textsubscript{3}N\textsubscript{4} nanobeam PhC cavities. (b) Normalized cavity reflection obtained by scanning a tunable laser around the cavity resonance. The Lorentzian fit yields a quality factor of Q $\sim$ 24,000. }
\label{Q_factor_results}
\end{figure}

For the quality factor measurement, a tunable laser was used and scanned around the TE fundamental mode, see Fig. \ref{laser_scan}. The experimentally measured quality factors for both devices are in the same order ranging from 10,000 to 24,000. Nevertheless, the dip transmission for device A is always larger as a result of the imperfection tolerance discussed above. It is worth noting that we have also fabricated devices on Si\textsubscript{3}N\textsubscript{4} membrane deposited via plasma-enhanced chemical-vapor deposition and found that quality factors always are always one order magnitude lower than previously mentioned. This is mainly due to the higher hydrogen impurity density which leads to higher optical absorption \cite{LPCVD_vs_PECVD_SiN}. Nevertheless, the  hydrogen impurity density can be decreased by annealing the nitride membrane at high temperatures \cite{PECVD_Annealing, PECVD_Annealing_2}.  

\begin{figure}
\centering     
\subfigure[]{\label{fiber_WG_Interface}\includegraphics[width=0.22\textwidth]{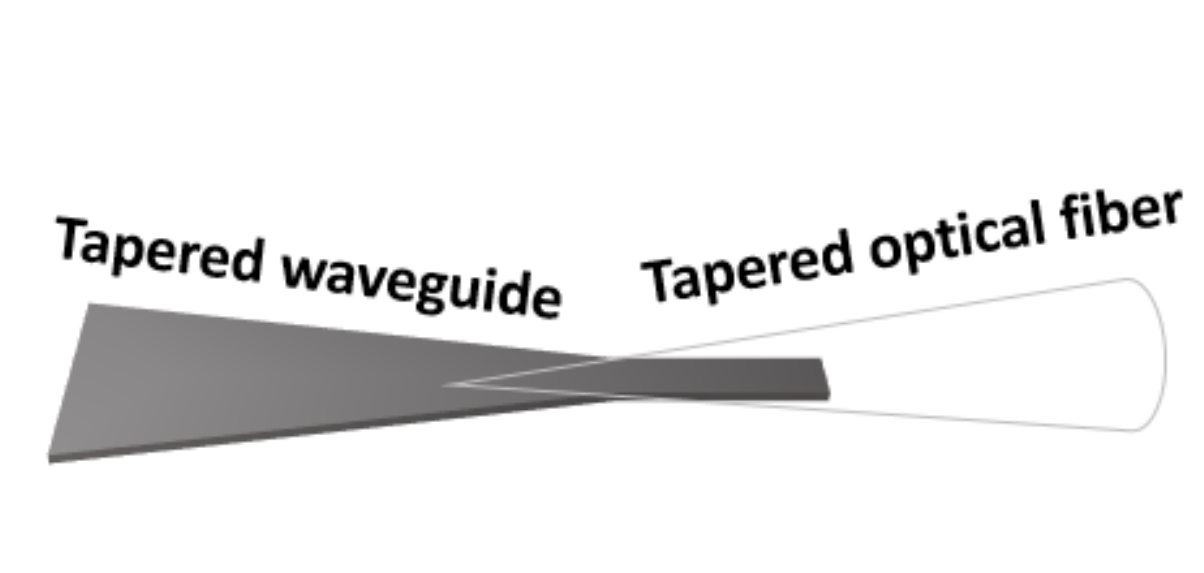}}
\subfigure[]{\label{neff}\includegraphics[width=0.2\textwidth]{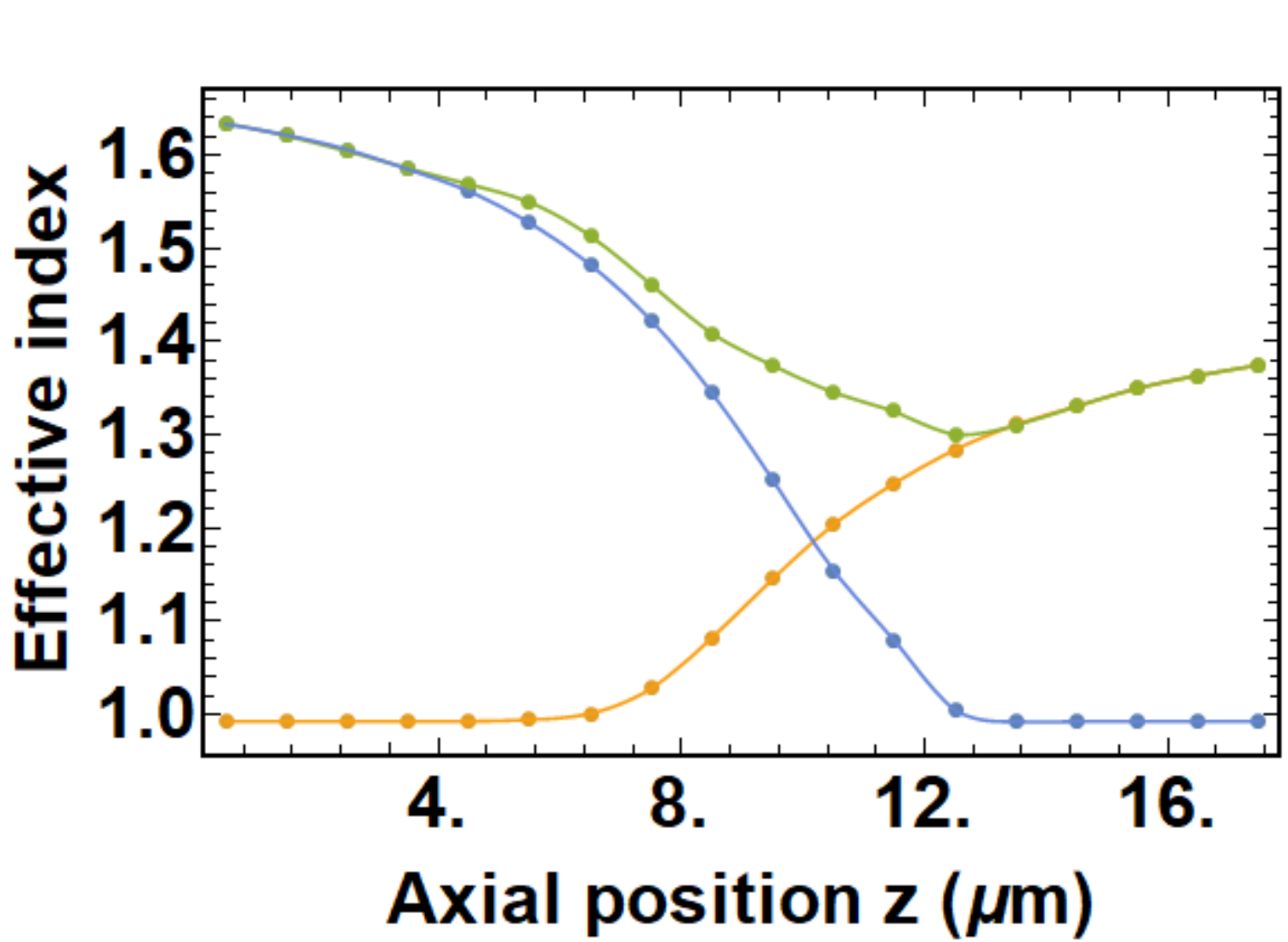}}
\subfigure[]{\label{E_Distribution}\includegraphics[width=0.45\textwidth]{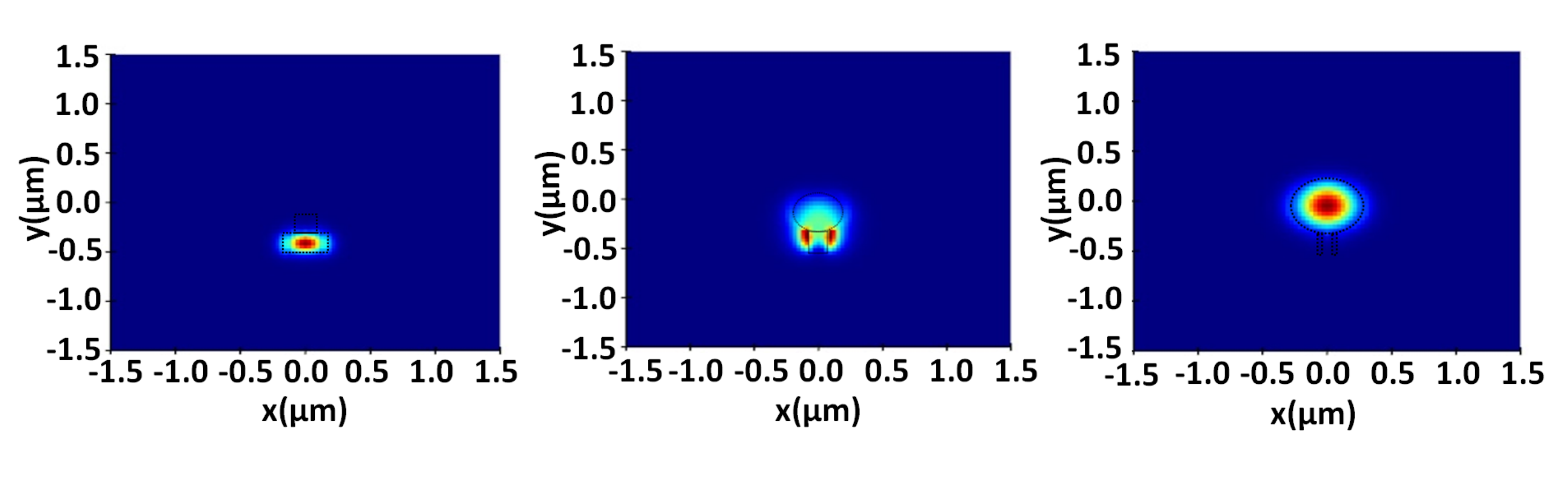}}
\subfigure[]{\label{width_plot}\includegraphics[width=0.2\textwidth]{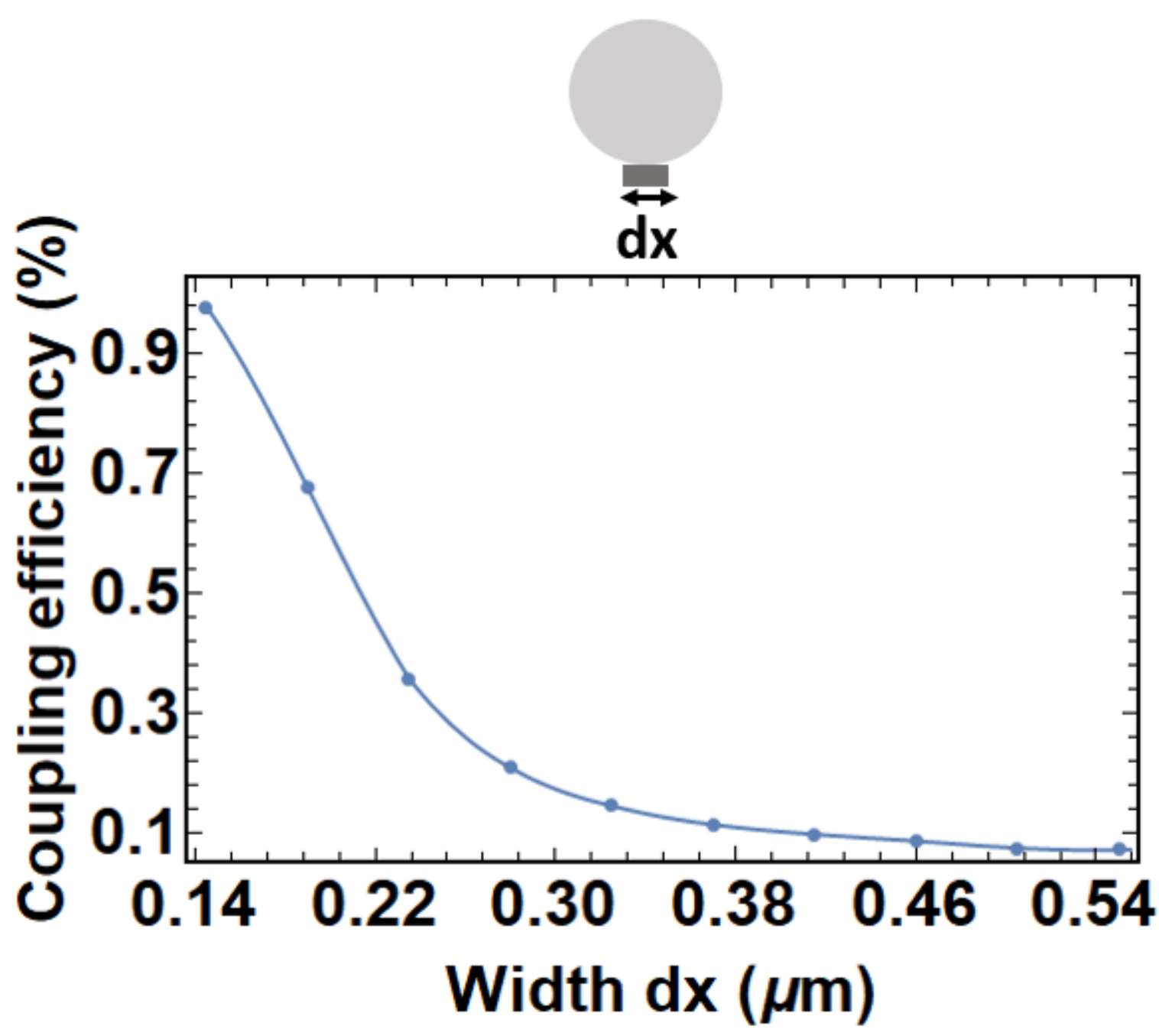}}
\subfigure[]{\label{overlapp_plot}\includegraphics[width=0.2\textwidth]{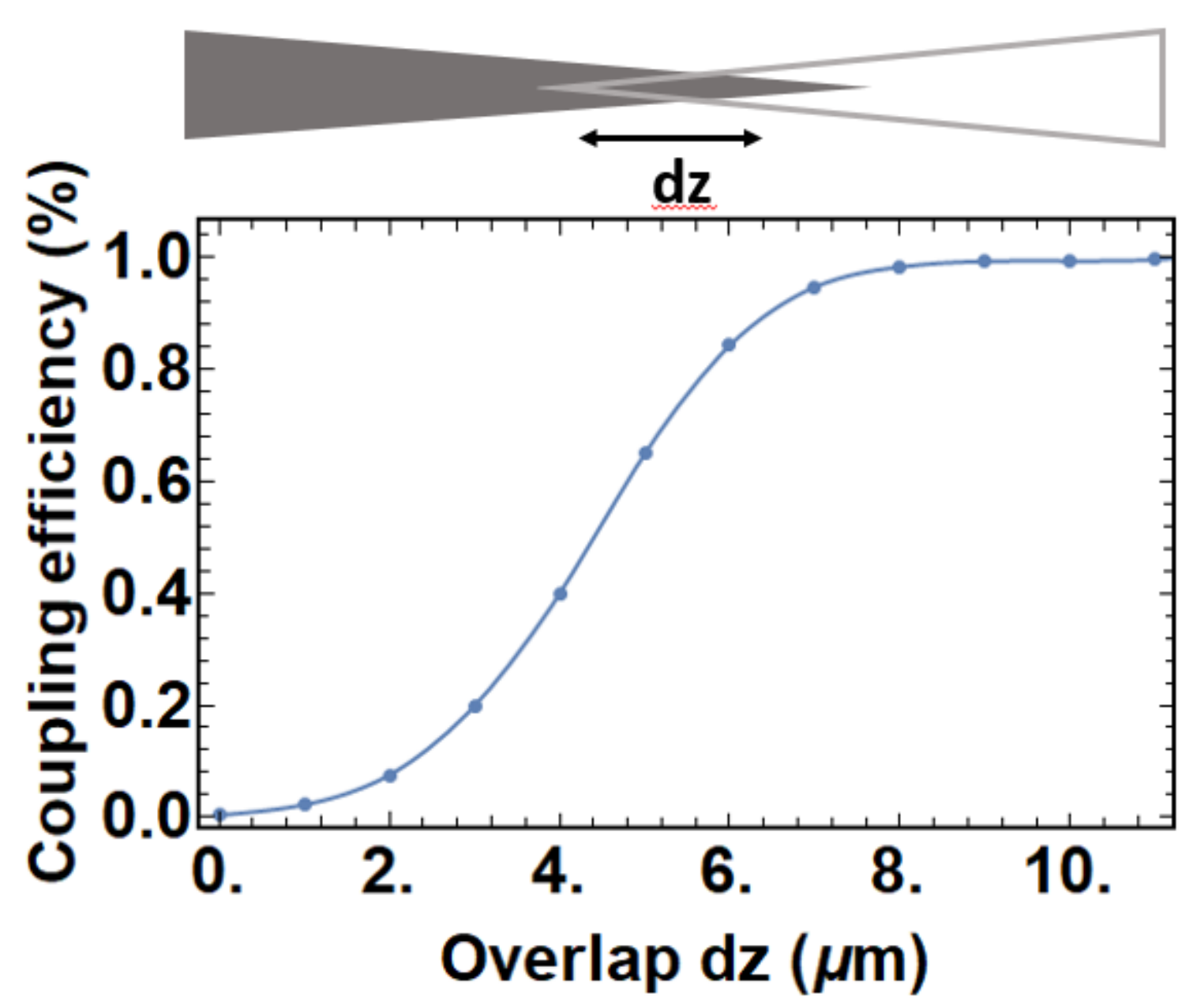}}
\caption{(a) Schematic of fiber-waveguide taper adiabatic coupling. (b) effective refractive index of the waveguide mode (blue), fiber mode (orange), and super-mode of the combined structure (green).  The opening angle of the fiber (waveguide) is of 3.5$^{\circ}$  (2.2$^{\circ}$). (c) Cross sections of $\textbar E \textbar ^2$ obtained from the FDTD simulation at different normal planes of the combined structure. (d) \& (e) Simulated coupling efficiencies for the power transmission from TE-polarized waveguide mode to the optical-fiber HE11 mode as a function of fiber-waveguide overlap and waveguide width, respectively}
\end{figure}

The light couples to the device using a single mode optical fiber tip that is coupled to the waveguide taper, see Fig. \ref{fiber_WG_Interface}.  This coupling technique is based on adiabatic transfer of an optical power between an optical fiber mode and a waveguide fundamental mode. The key idea is to change the effective refractive index of the target mode gently such that all the optical power are contained in the same mode profile. Therefore, the criterion of an adiabatic transfer is fulfilled if the change in the effective refractive index of the fundamental mode is less than the difference between effective refractive index of the fundamental mode and the nearest mode \cite{adiabatic_coupling_painter_2013}. This requires the tapering length to be longer than so-called beat length, the length over which the optical power could leak into a another mode. The beat length is given by $Z\textsubscript{b}=\lambda/(n\textsubscript{eff,1}- n\textsubscript{eff,2})$ \cite{adiabatic_coupling_ieee_91}, and it  corresponds to $Z\textsubscript{b}\backsim$ 2 $\mu$m for the geometry shown in Fig. \ref{neff}. The waveguide is stretched over 10 $\mu$m to ensure the stability of the optical fiber tip. In order to verify adiabatic mode transfer, FDTD simulations were performed in which a fundamental mode was launched at the beginning of the waveguide and monitored across the contact region, see Fig. \ref{E_Distribution}. Simulation results shows that $\sim 98 \%$ of the optical power has coupled to TE\textsubscript{11} optical fiber mode, while the other $2 \%$ leaks over the rectangular mechanical support that is connected to the chip. This result is in agreement to the side coupling approach aimed for detached devices \cite{fiber_optica_2015}. 

In order to characterize the influence of the rectangular mechanical support, we have calculated the coupling efficiency for various rectangle widths ranging from 1.4 $\mu$m to 0.5 $\mu$m. Figure \ref{width_plot} shows that the coupling efficiency drops exponentially as the rectangle width increases. This behavior is expected due to the swift change in the effective refractive index of the super-mode created in contact region. On the other hand, the length of the physical contact region in which an optical fiber tip overlaps with the waveguide taper is another crucial parameter. Figure \ref{overlapp_plot} shows that the coupling efficiency saturates after the overlapping region reaches 7 $\mu$m. This is relatively short compared to the biconical fiber taper technique previously reported \cite{adiabatic_coupling_painter_2013}.

\begin{figure}
\centering     

\subfigure[]{\label{fiber_etching}\includegraphics[width=0.2\textwidth]{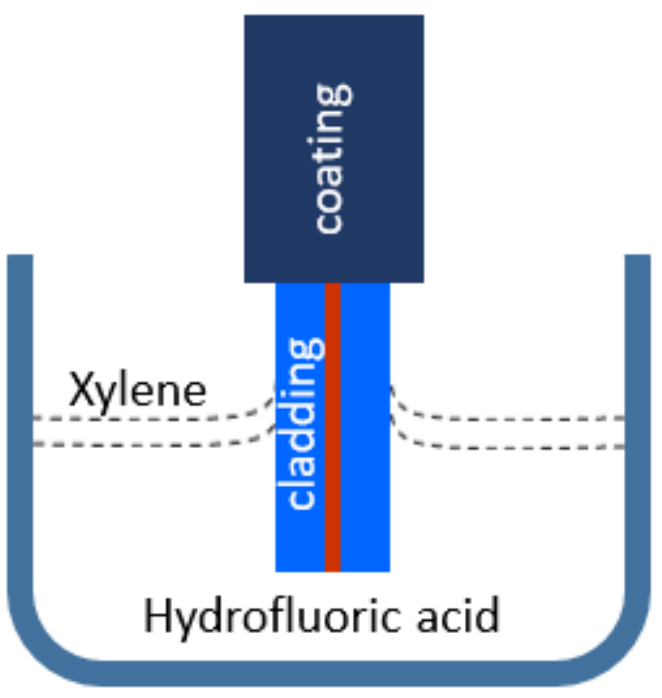}}
\subfigure[]{\label{fiber_SEM}\includegraphics[width=0.2\textwidth]{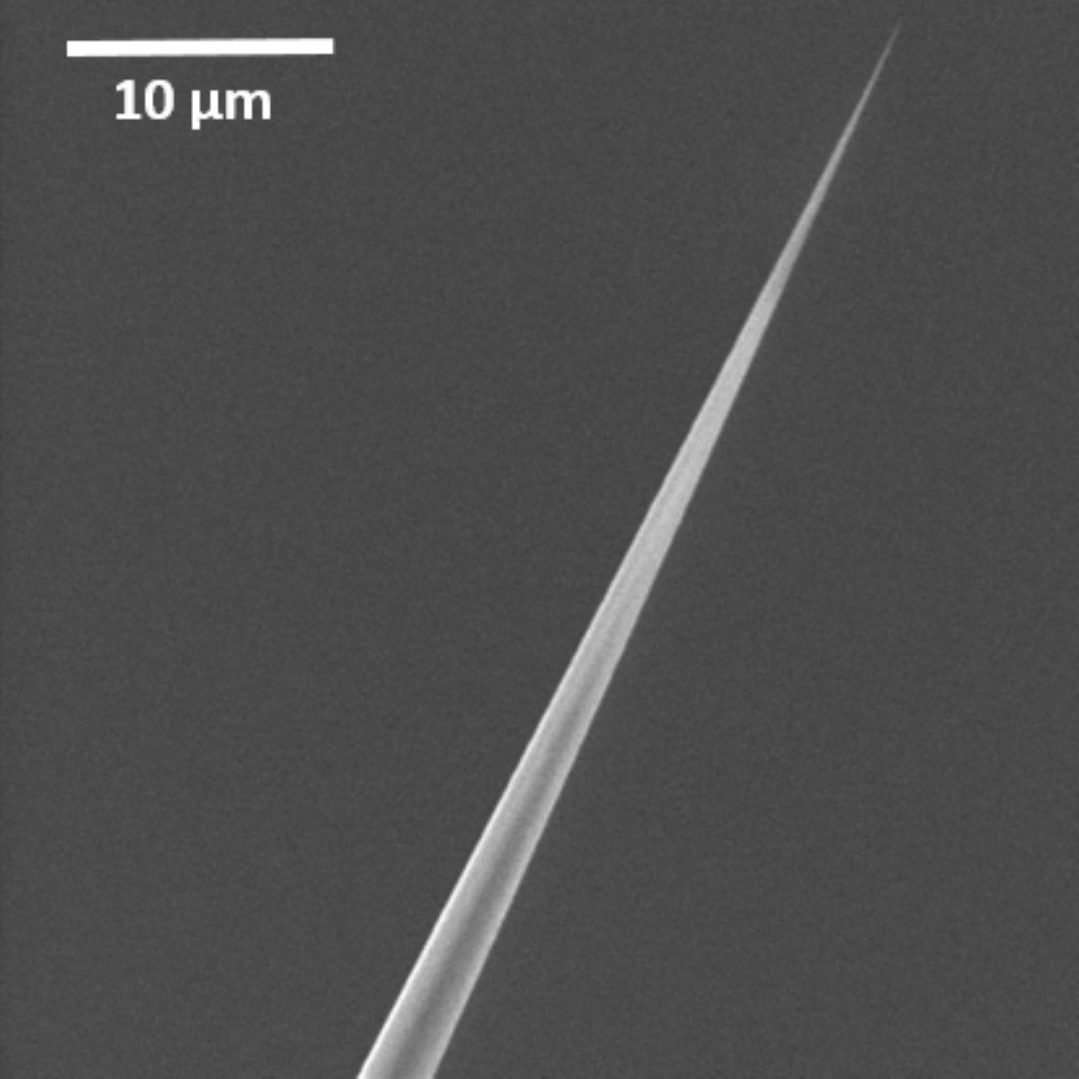}}
\subfigure[]{\label{coupling_setup}\includegraphics[width=0.33\textwidth]{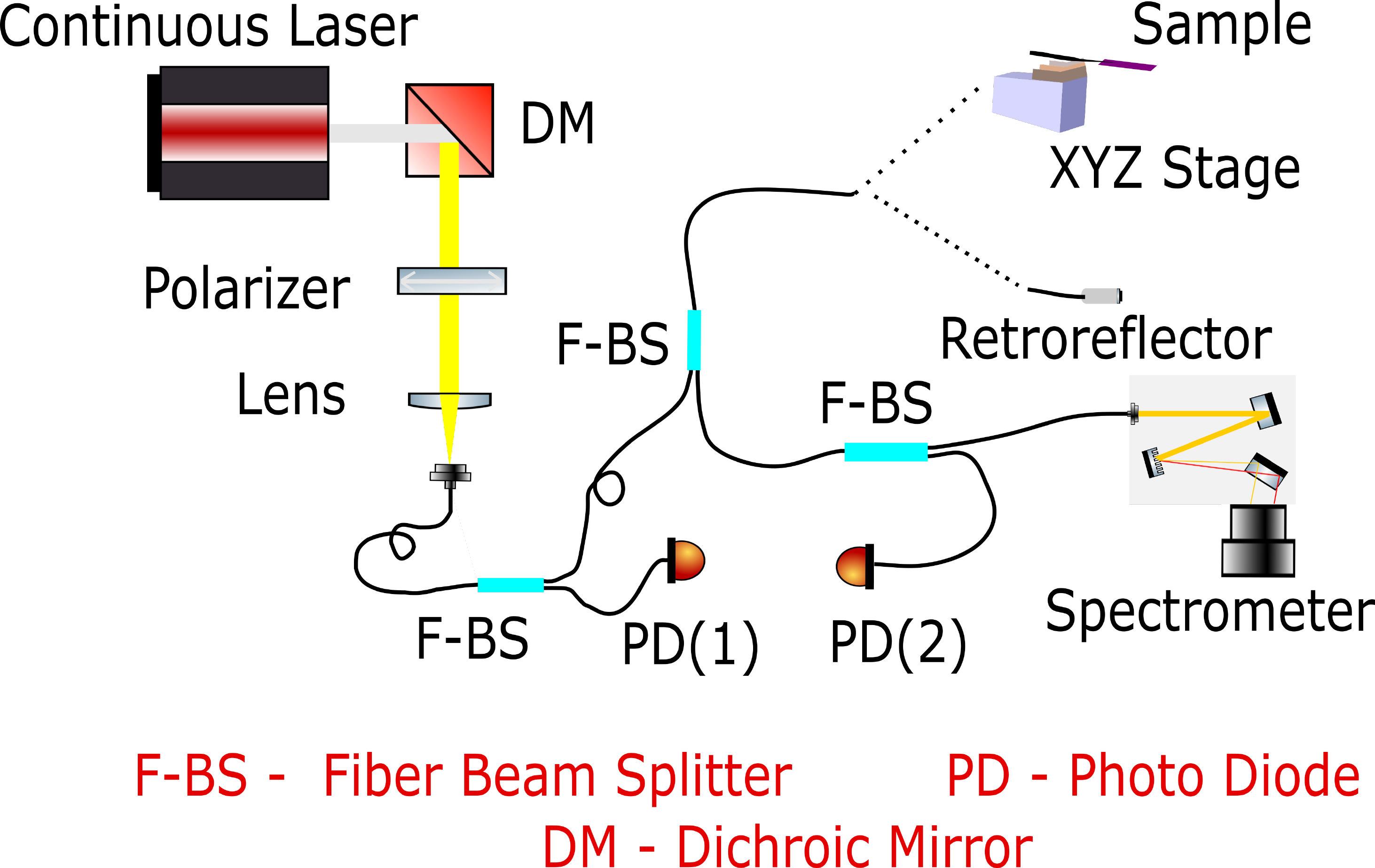}}
\subfigure[]{\label{CCD_Camera}\includegraphics[width=0.2\textwidth]{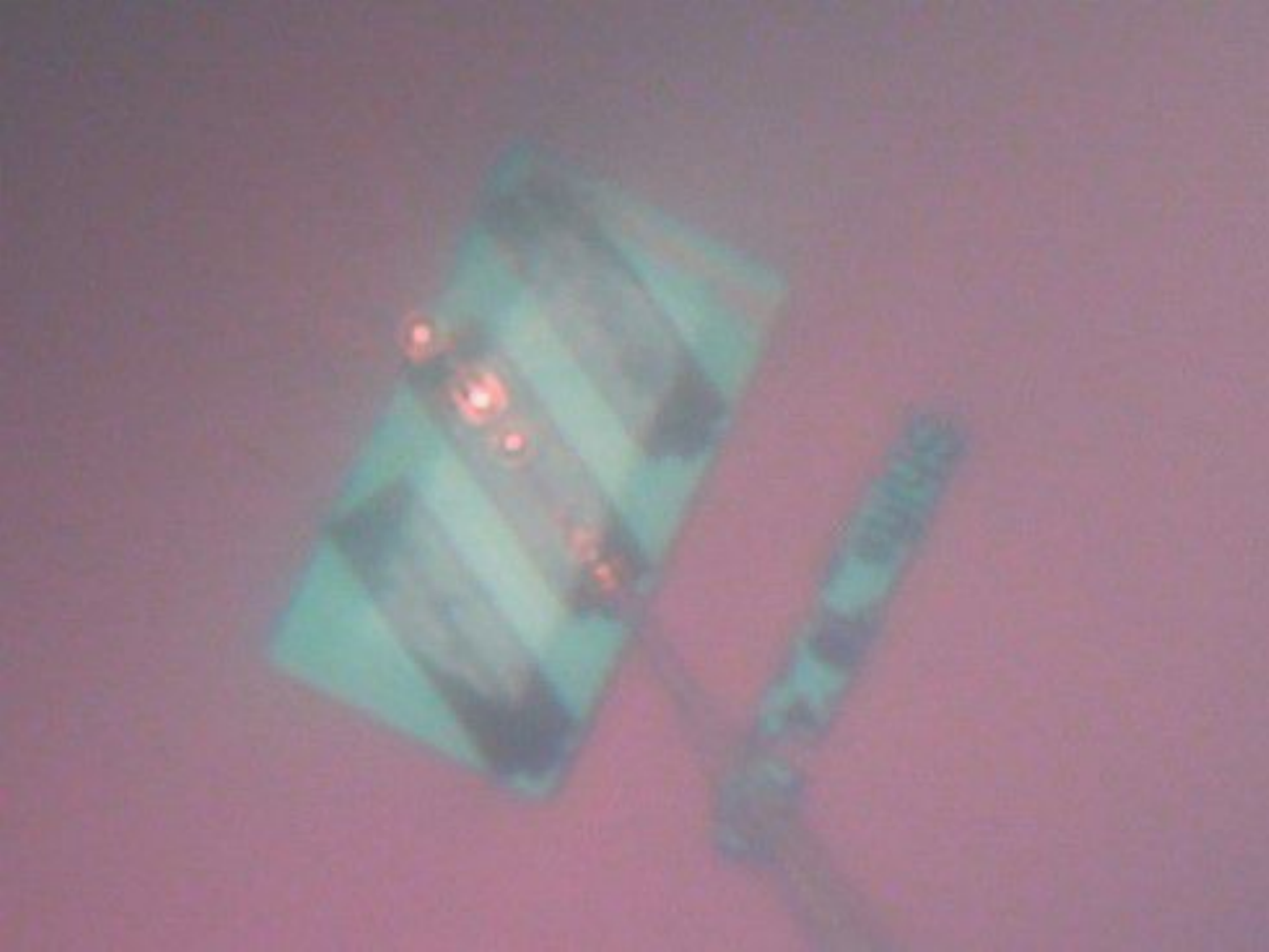}}
\subfigure[]{\label{retroreflector_spect}\includegraphics[width=0.2\textwidth]{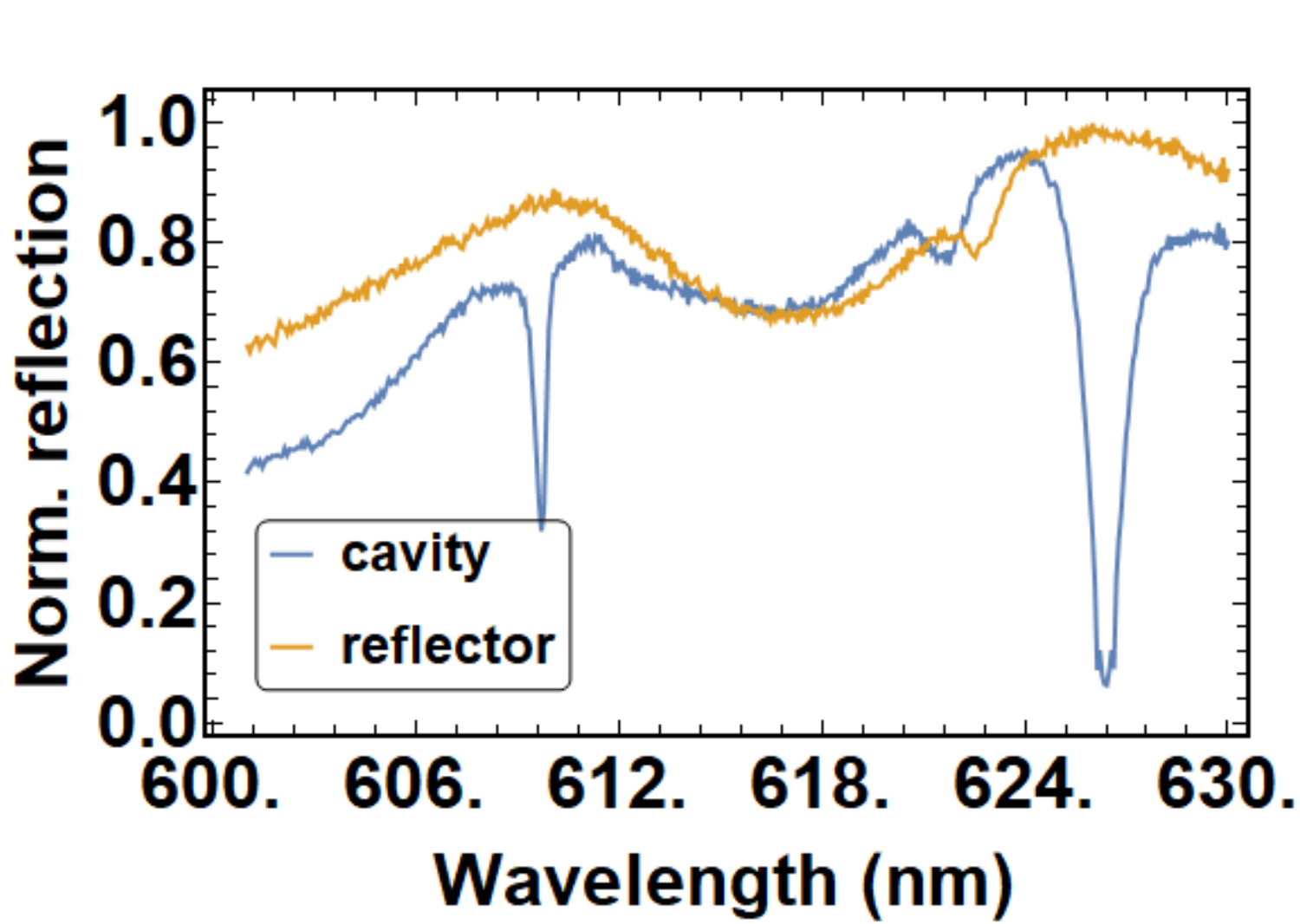}}
\caption{(a) Schematic of etching process of an optical fiber. (b) SEM image of an optical fiber tip after etching process was completed. (c) Schematic optical characterization of the coupling between a Si\textsubscript{3}N\textsubscript{4} nanobeam PhC cavity and an optical fiber. (d) Optical micrograph of a waveguide coupled to an optical fiber obtained from CCD camera. (e) Normalized broadband reflection spectrum of the cavity TE modes(blue) and a fiber-coupled retroreflector (yellow) obtained by a low resolution spectrometer.}
\end{figure}

The fabrication of an optical fiber tip is based on chemical wet etching approach in which a commercial single mode optical fiber (S405-XP, Thorlabs) is dunked in hydrofluoric acid (HF) covered by an organic solvent (e.g. xylene) to provide a oil-water interface \ref{fiber_etching}. The non-stripped fiber is clamped on a motorized stage (MT1-Z8, Thorlabs) to control the pulling speed via a motor controller (KDC101, Thorlabs). The fiber diameter continues to shrink gradually as it is pulled out of the HF solution. The height of the oil-water interface mainly depends on the fiber diameter and the surface tension difference between the acid and the organic solvent \cite{fiber_tip_fab_1999}. The key parameter in the fabrication process is the pulling speed that governs the angle of the optical fiber tip. The desired angle is dictated by the criterion of adiabatic transition of the fiber mode over the tapering region, similar to the waveguide taper discussed above. Since the cladding is etched away across the fiber tip, optical fiber cladding modes do not play any role in the coupling mechanism. This eases some restrictions in angles of the fiber tip  which is limited to $\theta < 5 ^{\circ}$ for an alternative technique that keeps the fiber cladding intact during the fabrication process  \cite{fiber_optica_2015}.   

The fabricated optical fiber tips were characterized by several measurements. We first measure the power transmission of an optical fiber tip using a custom setup where an objective of 0.6 numerical aperture is used to collect the power from the end of the optical fiber tip. The collected power is then normalized to the power obtained from a similar optical fiber with a clean core/cladding cross section. We routinely obtain high power transmission ($99\%$) even for optical fiber tips with relatively large angles ($10 ^{\circ}> \theta > 5 ^{\circ}$). This measurement is necessary to ensure the adiabatic transition of the fiber mode all over the optical fiber tip. However, we sometimes observe low power transmission due to the fabrication imperfections such as insufficient acid etching and inaccurate fiber adjustment of the clamp, and some remains of polymer coating. These remains can be removed in piranha cleaning prior to the acid etching.  

In the next measurement, we measure the coupling efficiency ($\mu \textsubscript{c}$) between optical fibers and suspended Si\textsubscript{3}N\textsubscript{4} cavities. The experiment is carried out as illustrated in Fig. \ref{coupling_setup}. A supercontinuum laser (SuperK EVo, NKT Photonics) is launched into a dichroic mirror and then polarized before coupled into the fiber network where two photodetectors (PD1 and PD2) are used to record the ratio between the incoming and reflected optical power ($\gamma$). The optical fiber is placed on top of a 3-axis stage controlled precisely by a piezo controller (MDT693B, Thorlabs). Next, we connect a retroreflector device to the incoming beam instead of the optical fiber. The resulting ratio of obtained from two photodetectors ($\zeta$) is used to calculate the coupling efficiency given by $\mu \textsubscript{c}^2 = \gamma/{\zeta \mu\textsubscript{Bragg}}$, where $\mu\textsubscript{Bragg}$ is the reflection of the cavity Bragg mirror which is assumed to be unity. For optical fibers with high power transmission, we routinely achieve a coupling efficiency $\mu \textsubscript{c} > 90\%$. The maximum coupling efficiency we have achieved is $\mu \textsubscript{c} = 96\%\pm 2\%$. The error bar reflects the fluctuations in the data collected by photodetectors which is averaged over a long period. It is worth noting that the coupling between an optical fiber and device B is always lower than device A. The underlying reason is attributed to the back reflection at the mirror-waveguide interface \cite{impedence_mismatch_2005}. Though we have appended transition holes to reduce the impedance mismatch in both devices, semicircular holes add more complexity at the transition region and require advanced tools for more precise adjustment. Nevertheless, we were able to achieved a coupling efficiency of $\mu \textsubscript{c} = 91\%\pm 2\%$ for device B. In qualitatively analysis, we compare broadband reflection spectrum obtained from the cavity and that obtained from retroreflector device. Figure \ref{retroreflector_spect} shows that the ratio between the broadband reflection spectra approaches unity, confirming the efficient fiber-waveguide coupling discussed above.

In summary, we have presented the design, fabrication, and characterization of nanobeam PhC cavities based on silicon nitride at visible wavelengths. We were able to demonstrate devices with quality factors higher than $10^4$ by scanning laser around the cavity resonance. The device offers promising a platform for strong coupling in solid-state cavity QED experiments. Specifically, we aim to integrate diamond color centers with silicon nitride photonics platform to realize an efficient light-matter interface. In addition to diamond color centers, our device can be used for other single photon emitters such as quantum dots \cite{colloidal_QD_2015} and layered 2D materials \cite{2D_Materials_hBN_2017}. Moreover, we have developed  a technique for coupling light from an optical fiber to on-chip suspended Si\textsubscript{3}N\textsubscript{4} nanobeam PhC cavities. Utilizing the coupling technique presented here, we demonstrated a coupling efficiency of $96\%$ between an optical fiber mode and a cavity mode. This work is useful for quantum optics applications, including long-distance quantum networking \cite{Q_internet_kimble} and optical quantum computing \cite{optical_Q_computing}.

\vspace{5mm} 

The authors would like to thank Denis Sukachev for valuable discussions. Financial support was provided by National Science Foundation (NSF) (PHY-1820930). Device fabrication is performed in part at the AggieFab Nanofabrication Facility at Texas A\&M University. The authors acknowledge the Texas A\&M University Brazos HPC cluster that contributed to the research reported here. The data that support the findings of this study is available from the corresponding author upon a reasonable request.

\bibliography{aipsamp}

\end{document}